\documentclass[aps,prstab,twocolumn,showpacs,superscriptaddress,groupedaddress]{revtex4}  
\usepackage{graphicx}  
\usepackage{dcolumn}   
\usepackage{natbib}
\usepackage[caption=false]{subfig}
\usepackage{bm}        
\usepackage{amssymb}   
\begin{document}



\title{Tapering studies for Terawatt level X-ray FELs with a superconducting undulator}
%
\affiliation{University of California, Los Angeles, California 90095, USA}
\affiliation{Stanford Linear Accelerator Center, Menlo Park, California 94025, USA}
\author{C.~Emma} \affiliation{University of California, Los Angeles, California 90095, USA}
\author{J.~Wu} \affiliation{Stanford Linear Accelerator Center, Menlo Park, California 94025, USA}
\author{P. Emma} \affiliation{Stanford Linear Accelerator Center, Menlo Park, California 94025, USA}
\author{Z. Huang} \affiliation{Stanford Linear Accelerator Center, Menlo Park, California 94025, USA}
\author{C.~Pellegrini} \affiliation{University of California, Los Angeles, California 90095, USA}\affiliation{Stanford Linear Accelerator Center, Menlo Park, California 94025, USA}
%
%
%
\vskip 0.25cm
\date{\today}

\begin{abstract}
We study the tapering optimization scheme for a short period, less than two cm, superconducting undulator, and show that it can generate 4 keV X-ray pulses with peak power in excess of 1 terawatt, using LCLS electron beam parameters. We study the effect of undulator module length relative to the FEL gain length for continous and step-wise taper profiles. For the optimal section length of 1.5m we study the evolution of the FEL process for two different superconducting technologies NbTi and Nb3Sn. We discuss the major factors limiting the maximum output power, particle detrapping around the saturation location and time dependent detrapping due to generation and amplification of sideband modes. 

\end{abstract}

\maketitle

\section{Introduction}
Much recent scientific effort has been devoted to studying the possibility of  using a tapered undulator \cite{KMR} to achieve terawatt level hard x-ray pulses in the next generation of Free Electron Lasers (FELs) \cite{wu2011simulation} \cite{HuangIPAC2012} \cite{makipac2014} \cite{geloniFEL2014} \cite{schneidmilleryurkovFEL2014}. The motivation for pursuing this goal comes primarily from the bioimaging community where a terawatt power coherent x-ray source would open the doors to single molecule imaging \cite{neutzenature} \cite{chapmannature} \cite{gaffneychapman}. Recent numerical work \cite{emmaprstab2014} \cite{PhysRevSTAB.Y.Jiao} shows that a self seeded hard X-ray FEL with LCLS-II like parameters in a 200-m permanent magnet undulator has the capability of reaching TW level pulses with the longitudinal and transverse coherence necessary for coherent X-ray imaging \cite{emmacoherenceFEL2014}. Spatial constraints, an increase in tunability and a longer machine lifetime have since led towards considering using superconducting technology for the undulator design rather than permanent magnets \cite{pemmaFEL2014}. 

In this study we present the results of tapering optimization for the cases of two different superconducting technologies, NbTi and Nb3Sn and assess the possibility of achieving TW levels of power in a 140 m undulator with periodic break sections. We study the impact of changing the undulator section length on the performance for both NbTi and Nb3Sn. We then present GENESIS simulation results for the NbTi case with the optimal choice of section length. Finally we discuss the problem of particle detrapping as a major source of performance degradation in tapered FELs, and propose some ideas to improve on current tapering designs by increasing the trapping and the extraction efficiency in the tapered section of the undulator. 

The physical system studied is a 140 m undulator composed of a 20 m SASE section followed by a a 120 m tapered undulator with sections 1 to 3 m in length. The SASE and tapered sections are separated by a self seeding chicane which delivers a 5 MW monochromatic seed at a photon energy of 4keV. We assume the undulator modules to be separated in all cases by 0.5 m break sections where we install the focusing quarupoles in a FODO configuration. The system parameters for both NbTi and Nb3Sn are described in table 1 where the beam energy is chosen in each case to generate 4keV photons. 

\begin{table}
\caption{\label{tab:table1}GENESIS Simulation Parameters}
\begin{ruledtabular}
\begin{tabular}{lcc}
Parameter Name& NbTi & Nb3Sn\\
\hline\hline
Electron Beam:& & \\
\\
Beam Energy $E_0$& 7.2 GeV &6.8 GeV\\
Energy Spread $\sigma_E$& 1.5 MeV&1.5 MeV\\
Peak Current $I_{pk}$& 4 kA & 4 kA\\
Normalized Emittances $\epsilon_{x,n}/\epsilon_{y,n}$& 0.4/0.4 $\mu$ m& 0.4/0.4$\mu$ m\\
Average $\beta$ function $\left\langle\beta \right\rangle$& 12 m& 12 m\\
\\
Undulator:& & \\
\\
Undulator Period $\lambda_w$& 20 mm& 18 mm\\
Undulator Parameter (RMS) $a_w$&  2.263& 2.263\\
Magnetic Gap $g$&  7.2 mm& 7.2 mm\\
Integrated Quad Field $B_q$&  4.5 T & 4.5 T\\
\\
Radiation:& & \\
\\
Photon Energy $E_{\gamma}$& 4keV & 4keV\\
Peak radiation power input $P_{seed}$& 5 MW  & 5MW\\
Seed laser size $\sigma_r$& 31 $\mu$ m  & 31 $\mu$ m\\
Rayleigh Range $Z_R$& 10 m & 10 m  \\
\\
FEL:& & \\
\\
FEL parameter $\rho $& 7.67 $\times 10^{-4}$ & 7.43 $\times 10^{-4}$ \\
FEL 3-D gain length $L_g^{3-D} $& 1.25 m &1.16 m \\
Fresnel Parameter $F_d $& 8 & 8.6\\
\end{tabular}
\end{ruledtabular}
\end{table}

The tapering optimization method used is the one described in Ref. \cite{PhysRevSTAB.Y.Jiao} with a transversely parabolic and longitudinally uniform electron beam distribution. The advantages of using a transversely parabolic beam as opposed to a Gaussian in a tapered X-FEL are described in detail in Ref. \cite{emmaprstab2014}. The quadrupole gradient is kept constant for simplicity and is set to achieve an average $\beta$ function of $\beta_{av}=$12 m throughout the undulator. The effect of varying the electron beam size by changing the quadrupole focusing strength will further increase the output power as discussed in Ref. \cite{PhysRevSTAB.Y.Jiao} however this option will not be examined in this work.

\section{Simulation Results}
\subsection{Section length study}

\begin{figure}[t]
\includegraphics[scale=0.31]{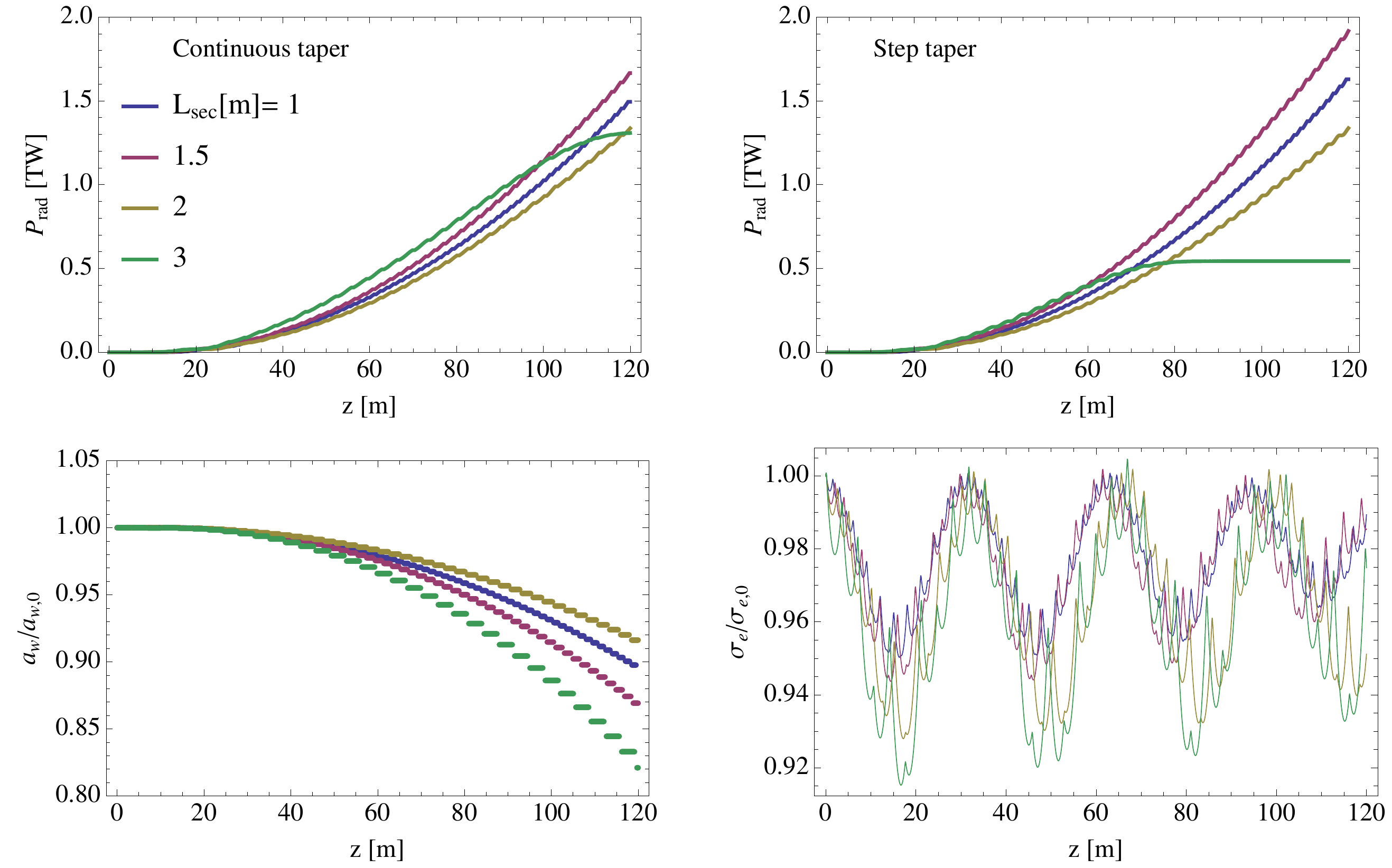}
\caption{Power evolution comparison for different undulator section lengths $L_{sec}$ separated by periodic 0.5 m breaks. The simulations  are time independent and use the Nb3Sn parameters listed in table 1 starting after the self seeding chicane (z=0). The transverse beam size is normalized to the input beam size $\sigma_{e,0}$ and the undulator parameter is normalized to the initial value $a_{w0}$.}
\end{figure}

In this section we investigate the effect of the undulator section length on the tapered FEL performance. Since the ideal tapering profile is a smoothly varying function of $z$ we expect that a longer section length limits the performance since shorter sections better approximate a continuous magnetic field profile. Furthermore the transverse beam envelope oscillations increase for larger section lengths $\Delta \beta^2/\beta_{av}^2=(\beta_{av} L)/(\beta_{av}^2-L^2)$ further degrading the FEL performance as shown in Fig. 1 for the continuous tapering case. We find that for the same average beta function $\beta_{av}$ as the section length becomes larger than the gain length the maximum output power is significantly reduced as illustrated in Fig. 1. Undulator sections of 1-2 m show a marginal improvement in the output power when going from the continuous to the step taper which can be attributed to the additional phase slippage of the electrons with respect to the ponderomotive wave in the step-wise case vs the continuous case. On the other hand in the 3 m case the maximum power decreases from 1.3 TW to 0.5 TW. This is a result of the more pronounced oscillations in the beam $\beta$ function and the phase mismatch between undulator sections as the discontinuity in magnetic field value is larger for longer sections. The 1 m undulator sections are thus optimal for minimizing oscillations in the beam size and approximating a smooth taper profile however the filling factor for such short sections limits the interaction length and overall extraction efficiency. Thus we determine that to obtain a reasonable filling factor and large output power a 1.5 m section length is the optimal choice for the undulator design.

\subsection{FEL evolution with 1.5 m undulator sections}

\begin{figure}
\includegraphics[scale=0.3]{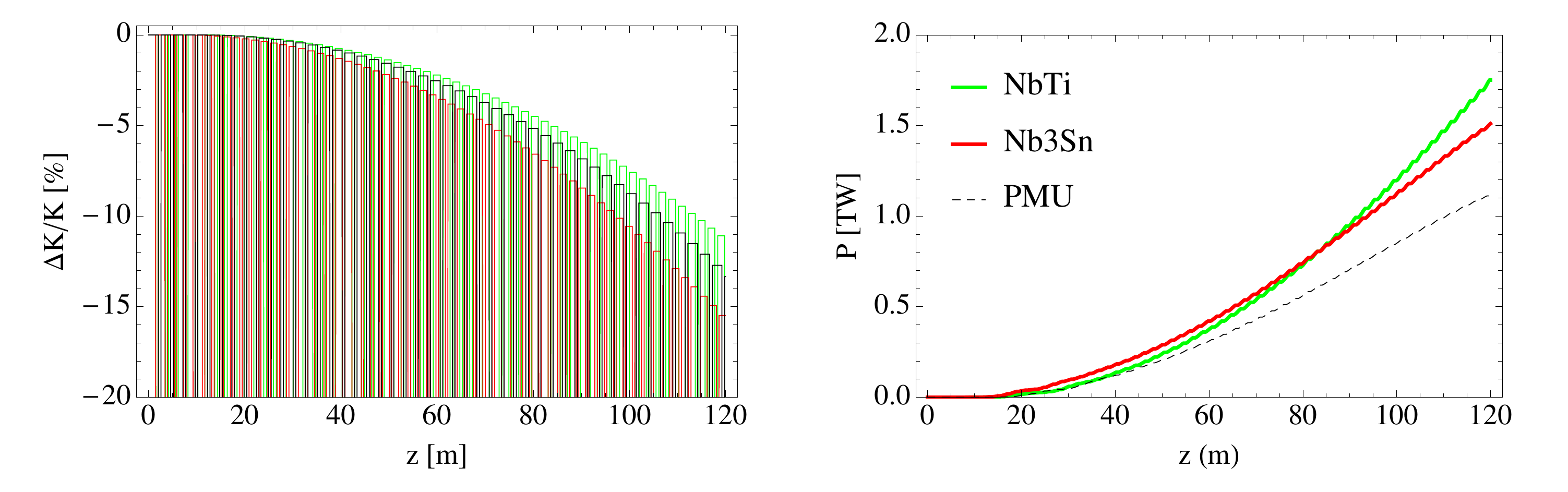}
\caption{Time independent GENESIS simulation results for the optimized taper profile and evolution of the FEL radiation power for NbTi and Nb3Sn superconducting undulators. We also plot for comparison the case of a permanent maget undulator of 2.6 cm period and undulator parameter $a_{w0}=$ 1.77 as discussed in the text.}
\end{figure}

For our optimal choice of section length time, independent tapering optimizations produce the magnetic field profiles displayed in Fig. 2. In both cases the optimal functional form of the tapering law is very close to quadratic $a_w(z)\sim a_{w0}\left[1-c(z-z_0)^2\right]$ with varying taper strengths $c$ and a start location around $z_0\sim 10$ m after the self-seeding chicane. The quadratic scaling can be understood from the 1-D theory of tapered FELs \cite{KMR}. The change in energy for a resonant electron is determined by a $z$ dependent poderomotive gradient:

\begin{equation}
\frac{d\gamma_r^2}{dz}=-\frac{eE(z)a_w(z)}{mc^2}\sin \Psi_R 
\label{1dequations}
\end{equation}

where for simplicity we assume the resonant phase $\Psi_R$ to be constant, $E(z)$ is the magnitude of the growing radiation field and $a_w(z)$ is the RMS magnitude of the tapered magnetic field. The FEL will radiate at a wavelength $\lambda_s$ if the following $z$ dependent resonance condition is satisfied:

\begin{equation}
\gamma_r^2(z)=\frac{\lambda_w}{2\lambda_s} \left(1+a_w(z)^2\right)
\end{equation}

If we assume that the radiative process in the tapered section of the undulator is mostly due to coherent emission, with the electric field evolving according to $E(z)\propto z$, to satisfy simultaneously Eq. 1-2 the magnetic field $a_w(z)$ must decrease quadratically with $z$. With the optimal taper profiles for NbTi and Nb3Sn we obtain extraction efficiencies of $6.08 \%$ and $5.89 \%$ respectively, and output powers over a factor of 75 larger than the saturation power in both cases (see Fig. 2). Applying the same tapering optimization for a Permanent Magnet Undulator (PMU) with a 26 mm period, beam energy of 6.7 GeV and an RMS undulator parameter $a_{w0}$=1.77 we report a $\sim$ 60 $\%$ reduction in output power compared to the superconducting case with $P_{rad}^{PMU}=1.1$ TW at the undulator exit.  As mentioned previously this is one of the motivations for moving from permanent magnets to superconducting technology in the design of Terawatt level X-FELs such as LCLS-II. We also note that for the superconducting case a reduction in the average $\beta$ function will further increase the efficiency while this cannot be done in the permanent magnet case.

\begin{figure}[t]
\includegraphics[scale=0.2]{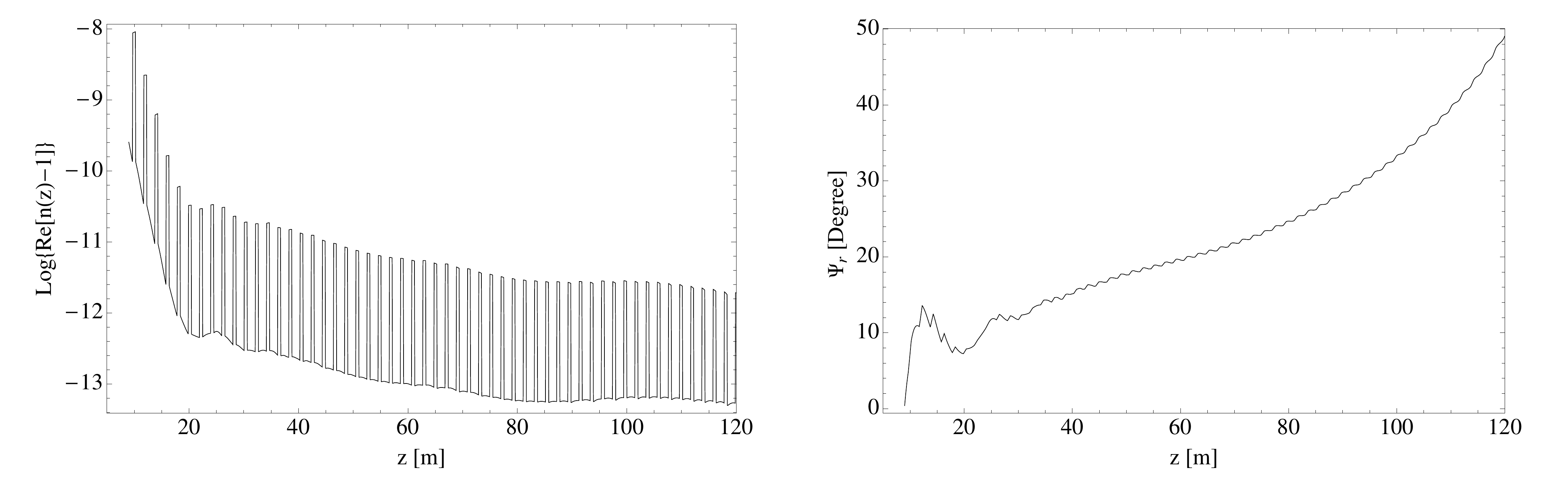}
\caption{Real part of the electron beam refractive index (left) and on axis resonant phase (right) for NbTi superconducting undulator parameters from time independent GENESIS simulation.}
\end{figure}

\subsection{Limits on optimization due to parasitic effects}
The FEL process is dominated by the effects of refractive guiding and particle detrapping in the tapered section of the undulator. In Fig. 3 we show the evolution of the real part of the refractive index \cite{PhysRevA.24.1436} (in the resonant particle approximation) and the resonant phase given by the expressions:

\begin{equation}
\Re(n-1)=\frac{\omega_{p0}^2}{\omega_s^2} \frac{r_{b,0}^2}{r_{b}^2} \frac{a_w}{2|a_s|}[JJ]\left\langle\frac{\cos \Psi_R}{\gamma_R}\right\rangle
\end{equation} 
\begin{equation}
\sin\Psi_R(z)= \chi\frac{|a_w'(z)|}{E(z)}
\end{equation}

where $\chi=(2*me*c^2/e)(\lambda_w/2\lambda_s)(1/\sqrt{2}{[JJ]})$ is a constant independent of $z$ and the other symbols have their usual meaning. From these expressions we see that if we want to maintain strong optical guiding, the bunching must be preserved during the tapered section of the undulator. However as the radiation field grows the effect of optical guiding will decrease as $1/|a_s|$ inducing a self-limiting mechanism on the growth of the field \cite{fawleynim1996}. Furthermore, in order to provide the largest ponderomotive gradient to induce the greatest energy loss in the electrons, it is desirable to increase the resonant phase throughout the tapered section. Again here we find a self-limiting mechanism which limits how much one should increase $\Psi_R$, since a larger $\Psi_R$ results in a smaller ponderomotive bucket creating a tradeoff between the ponderomotive potential strength and the number of trapped electrons \cite{KMR}. From Fig. 3 and 4 we see that as the resonant phase increases from zero to $\psi_r=20^o$ from $z$=10-50 m, the phase shift in the ponderomotive bucket causes $\sim 20 \%$ of the particles to detrap. This can be mitigated by introducing phase shifters in the break sections between 10-50 m and is an effect which will be examined quantitatively in future studies. 

\begin{figure}[t]
\includegraphics[scale=0.32]{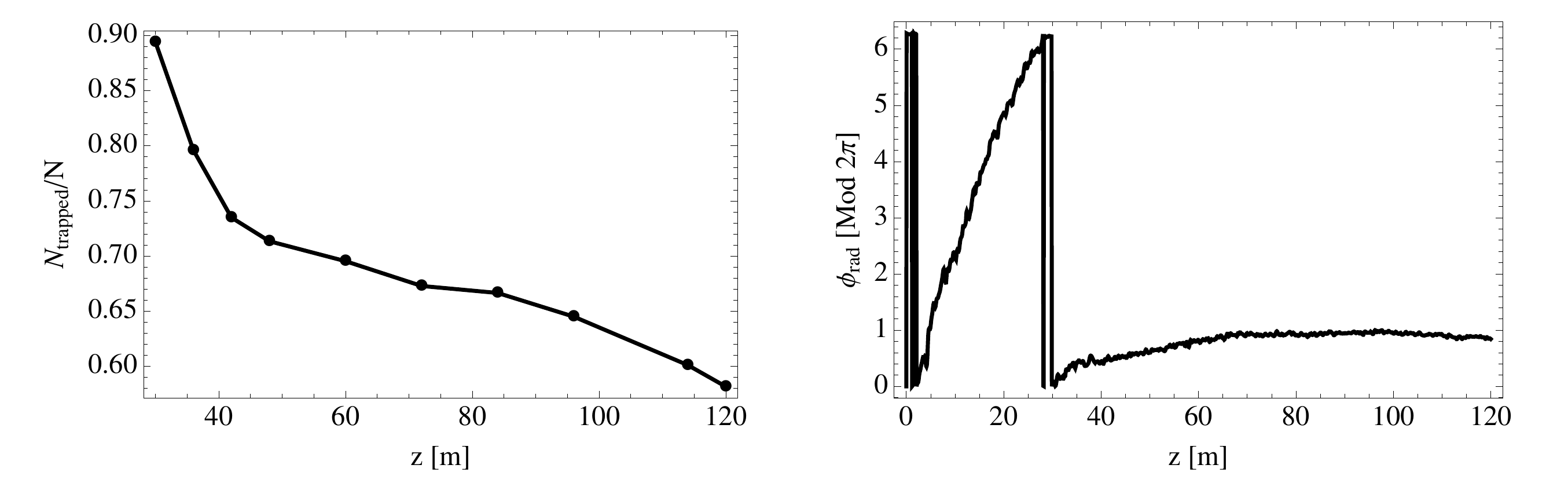}
\caption{Radiation Phase (on axis) and trapping fraction from time independent GENESIS simulations of NbTi. The area after 40 m is when the gain is so low that the phase velocity of light is almost $c$ thus the radiation phase is constant}
\end{figure}

\subsection{Time dependent effects. Sideband growth and sideband induced detrapping}
As has been pointed out in previous work \cite{TESSA} \citep{PhysRevSTAB.Y.Jiao} \cite{emmaprstab2014}, the performance of tapered hard X-ray FELs is also limited by particle detrapping due to time dependent effects. Shot-noise fluctuations in the electron beam current induce modulations in the radiation beam longitudinal profile. This can cause particle detrapping from the top and bottom of the ponderomotive bucket as the modulated radiation field slips through the beam and the particles experience fluctuations in the bucket height. Furthermore, the resonant interaction between the FEL primary wave $k_0$ and the electron synchrotron motion drives a sideband instability which amplifies parasitic modes at wavenumbers $k_0 \pm \Omega_s$ where $\Omega_s$ is the synchrotron wavenumber:

\begin{equation}
\Omega_s^2(z)=\frac{eE(z)a_w(z)}{m_ec^2k_w}
\end{equation}

The growth of the sidebands not only increases the bandwidth of the FEL signal but also induces further particle detrapping \cite{sidebandsdavidson} \cite{sidebandstang}. \newline

\begin{figure}[t]
\includegraphics[scale=0.3]{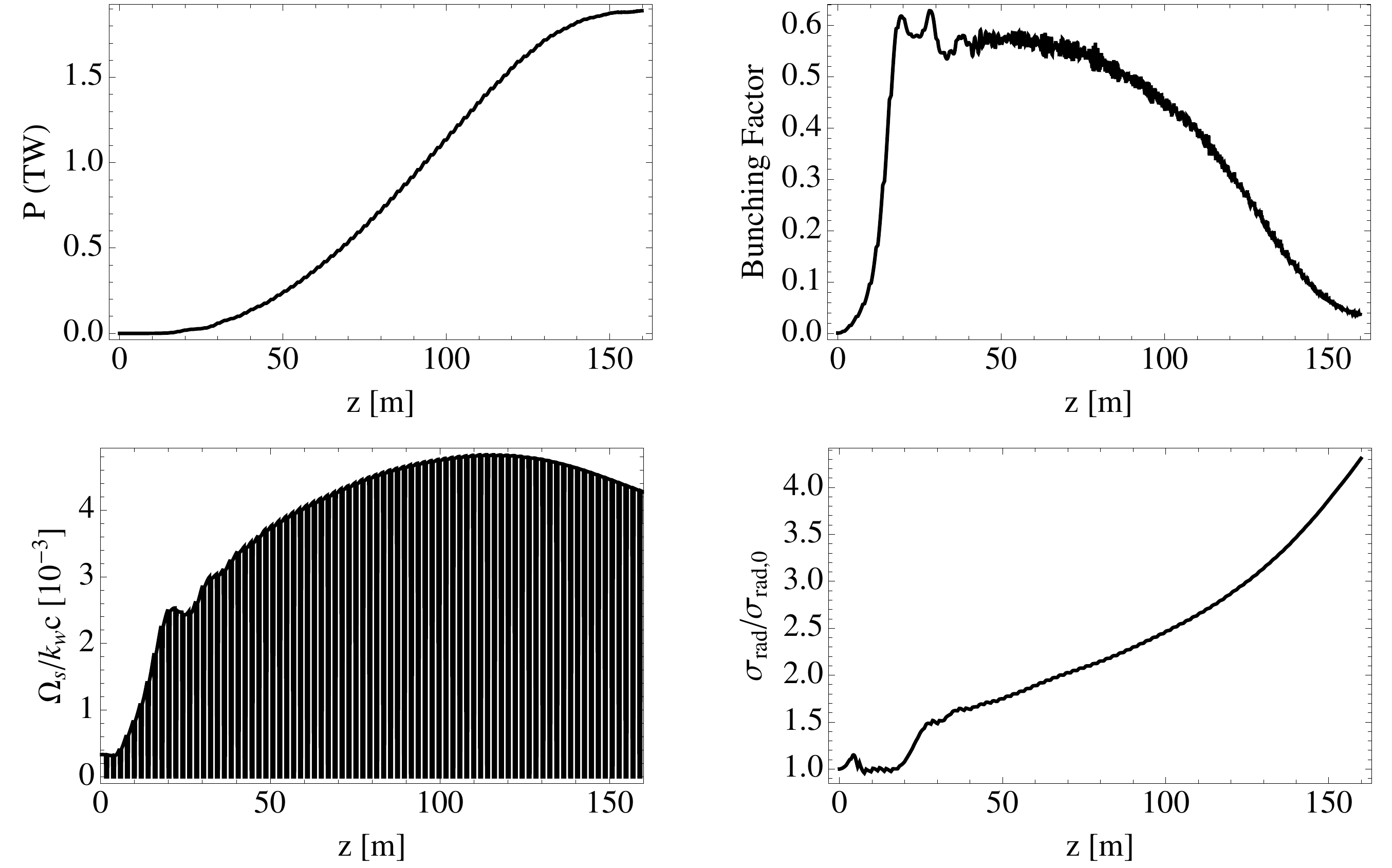}
\label{tdpresults}
\caption{Time dependent GENESIS simulation results for the NbTi case parameters in Table 1. The beam is 37.5 fs long with a flattop longitudinal profile.}
\end{figure}

\begin{figure}
\includegraphics[scale=0.5]{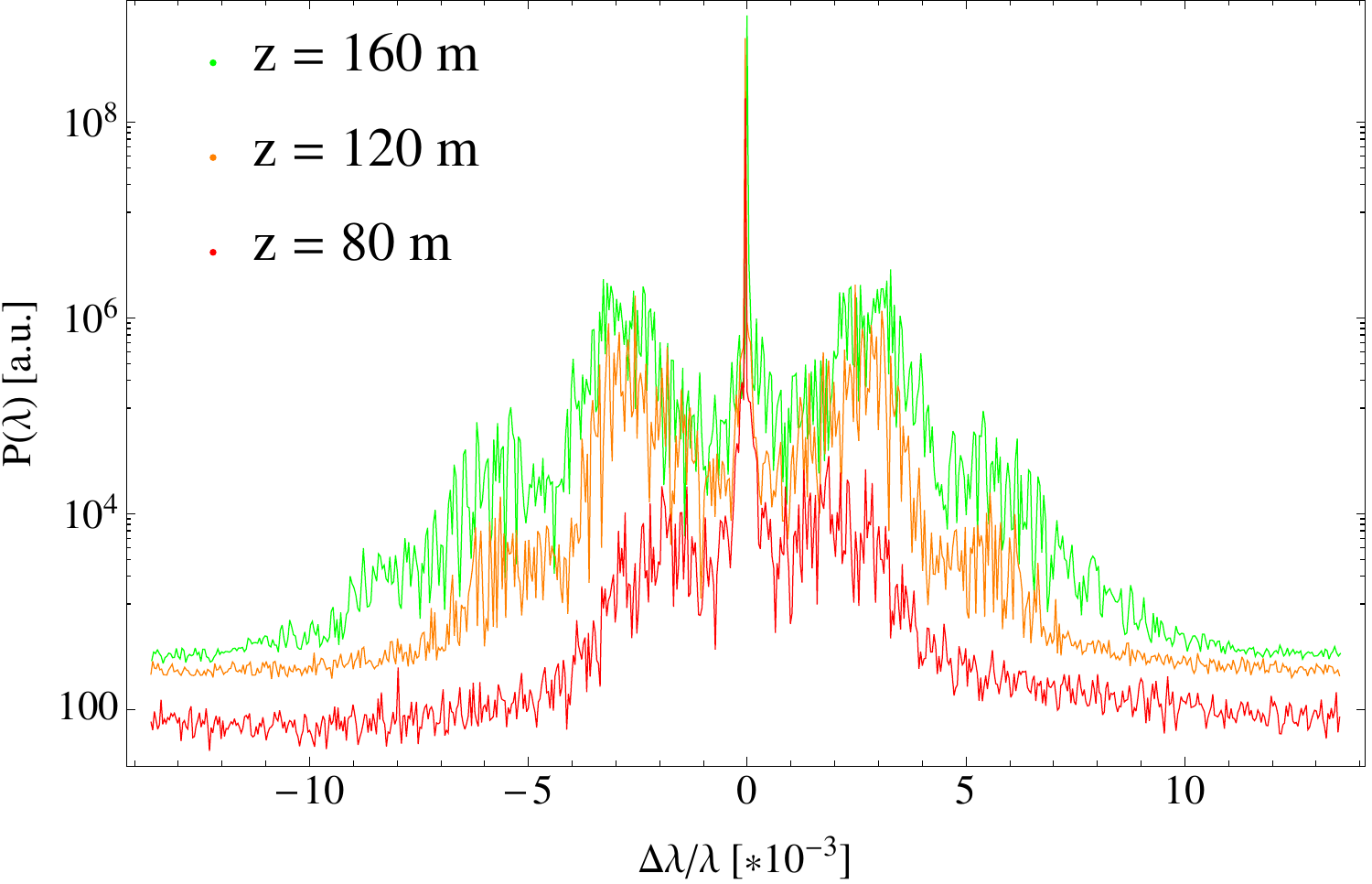}
\caption{Growth of the FEL synchrotron sidebands at different locations in the undulator. The growth of the sidebands intensified in the later section of the undulator where the synchrotron frequency is changing less rapidly (see Fig. 5).}
\end{figure}

Time dependent effects are analysed by simulating the beam parameters in table I for a longitudinally uniform bunch of 37.5 fs. The simulation results are shown in Fig. 5 for a longer undulator of 160 m. The simulation was done for a longer undulator to emphasize the peak power saturation in the final sections of the tapered undulator. From Fig. 5 we can see that there is a difference in peak output power (averaged over time) compared with the time independent results, namely 1.78 TW in the time independent case compared to 1.55 TW after 120 m time dependent. The drop can be attributed to the growth of the sidebands shown in a plot of the spectrum at various $z$ locations (see Fig. 6). This growth can be mitigated by changing the synchrotron frequency along the tapered section of the undulator \cite{Pillasidebands} \cite{sharpyusidebands}. In the final section of the undulator ($z>$100m) the variation in the synchrotron frequency is small due to the saturation of the electric field. It might be possible to compensate for this effect by introducing a cubic or quartic term to the taper profile. This option will also be explored in the future. 

\section{Conclusion}

We have presented a numerical study of tapering optimization for a self seeded hard X-ray FEL with superconducting undulator parameters. This work has demonstrated numerically that peak power levels over 1 TW can be achieved in a 120 m undulator with break sections and an optimized taper profile for two separate superconducting technologies: NbTi and Nb3Sn. We demonstrated a 60$\%$ increase in output power for the superconducting technology when compared with a permanent magnet design. We have also outlined the effects which limit the growth of the power in the tapered undulator and have found that particle detrapping is the main obstacle to achieving larger extraction efficiencies. The two causes of particle detrapping we have discussed are due to an increase of the resonant phase after the exponential gain regime and sideband induced detrapping due to time dependent effects. To mitigate these two effects we propose to study applying phase shifters at the location of initial saturation and introducing a faster term in the taper profile to reduce the sideband growth by changing the synchrotron frequency faster in the final sections of the undulator. 

\section*{ACKNOWLEDGMENT}
The authors would like to thank Dr. G. Marcus and J. Duris for useful discussions. This work was supported by U.S. D.O.E. under Grant No. DE-SC0009983.

\bibliographystyle{ieeetr}
\bibliography{prstabreferences.bib}

\end{document}